# Nonreciprocal spontaneous parametric process

Changbiao Li[1,*], Jiaqi Yuan[1,*], Ruidong He[1], Jiawei Yu[1], Yanpeng Zhang[1], Min Xiao[2,3], Keyu Xia[2,†], Zhaoyang Zhang[1,‡]

[1]*Key Laboratory for Physical Electronics and Devices of the Ministry of Education & Shaanxi Key Lab of Information Photonic Technique, School of Electronic and Information Engineering, Faculty of Electronic and Information Engineering, Xi'an Jiaotong University, Xi'an, 710049, China*

[2]*National Laboratory of Solid State Microstructures and School of Physics, Nanjing University, Nanjing 210093, China*

[3]*Department of Physics, University of Arkansas, Fayetteville, Arkansas, 72701, USA*

[*]These authors contributed equally to this work.

Corresponding authors: [†]keyu.xia@nju.edu.cn, [‡]zhyzhang@xjtu.edu.cn

Mediated by the interaction with quantum vacuum fields, a laser field propagating in a nonlinear optical medium can generate new light fields via spontaneous parametric process. Such process is inherent independent of the propagation direction of light and reciprocal thus far, due to the direction-independent field-vacuum interaction. In this work, we experimentally demonstrate a nonreciprocal spontaneous parametric four-wave mixing process in sodium atomic vapors with dispersive nonlinearity and further broadband optical isolation by unidirectionally coupling the probe field to an auxiliary quantum vacuum field in another four-wave mixing process. Thanks to the broad bandwidth of the spontaneous parametric process, in combination with the Doppler and power-induced broadening of atomic energy levels, we achieve optical isolation with a bandwidth larger than 100 GHz for isolation ratio ≥25 dB. Considering that both spontaneous parametric processes and wave mixing in nonlinear medium have been realized in diverse on-chip photonic platforms, our work paves the way for integrated broadband optical isolations and thus can boost scalability and function of photonic chips.

A basic concept in classical electromagnetic theory based on the Maxwell's equations is that the electromagnetic field vanishes in a "null" closing space in which internal sources and external driving are absent. In stark contrast, quantum optics predicts a non-zero quantum fluctuation, known as the quantum vacuum field (QVF), even in a null closing space. This QVF is one of deepest fundamental properties of nature. It has been directly observed with a superconducting atom [1]. QVFs play critically important roles for understanding many fundamental quantum processes such as spontaneous decay of atoms [2], Casimir effect [3, 4], and cavity quantum electrodynamics [5], to name a few. Very recently, quantum vacuum fluctuation is employed to explain the origin of chiral molecules [6]. Beside fundamental physics, engineering QVFs has already demonstrated vacuum induced transparency [7], single-photon transistor [8], sensing enhancement [9] and even manipulation of matter [10]. Importantly, QVFs can efficiently trigger the generation of new paired electromagnetic fields from a coherent probe laser field via the spontaneous parametric process in a nonlinear optical medium [11], when the involved fields simultaneously meet the phase-matching condition (PMC) and energy conservation. This process acts as the basis of heralded single-photon sources for quantum information technologies and modern optics. By far, spontaneous parametric processes are reciprocal because of the time-reversal symmetry and spatial inversion symmetry of quantum fluctuation. The current work theoretically proposes and experimentally demonstrates a nonreciprocal spontaneous parametric four-wave mixing (FWM) process for realizing broadband optical isolation.

Nonreciprocal optical devices (NORDs) breaking the Lorentz reciprocity can perform optical isolation by enforcing one-way propagation of light. Such devices play a vital role in laser protection, optical and integrated photonic technologies [12], and even quantum information processing [13-17]. They are conventionally realized with the magneto-optical effect [18]. Such magneto-optical devices have the unique merit of broad bandwidth, yet their compact and integrated implementation as a whole component remains an open challenge.

Magnetic-free NRODs are therefore developed to tackle this challenging problem in footprint. Mechanisms in the classical regime include optical nonlinearity [19-29], phonon-photon directional coupling [30-35], the Sagnac effect in spinning resonators [36,37], the macroscopic Doppler effect of unidirectionally moving Bragg optical lattice [38-40], and directional phase matching in pumped wave mixing and parametric process [41,42]. Besides, quantum systems have also demonstrated a great success in novel non-magnetic NRODs by engineering reservoir of resonators [43], exploiting quantum nonlinearity [44], chiral interaction of atoms and photons with the spin-momentum locking [45-51], the susceptibility-momentum locking [52-59] and unidirectional quantum squeezing [60]. In comparison with the magneto-optical counterparts, the practical applications of these magnetic-free approaches are often limited by their narrow bandwidth due to the requirement of high-quality resonators, or long lifetime of adopted systems. Among aforementioned approaches, all-optical NRODs based on wave conversion attracts intensive attention because they are inherently compatible with integrated photonic technologies. Nonetheless, the precise PMC and energy conservation impose a strong constrain on their applications in bandwidth. Benefiting from the broadband PMC, a nonreciprocal spontaneous parametric process exhibits an essential advantage as it has the potential to overcome the limitation of bandwidth in optical isolation.

In this paper, a proof-of-concept experiment is performed to show that a spontaneous parametric FWM (SPFWM) process driven by a weak probe laser beam can be nonreciprocal by inducing direction-dependent probe-vacuum interaction, leading to chiral energy conservation. By realizing such nonreciprocal SPFWM process in hot sodium (Na) atoms, we obtain optical isolation over 100 GHz bandwidth (corresponding to ~0.12 nm @ 589 nm) for isolation ratio > 25 dB, and a maximum ratio larger than 30 dB. In thermal atomic ensembles, the inevitable Doppler and power-induced broadening of the decoherence rates of associated atomic energy levels are usually detrimental due to their suppression on the efficiency of desired atomic coherence effects. Counterintuitively, such broadening effects in our experiment act as the useful resource to extend the bandwidth of the dispersive nonlinear response and further enable the broadband optical isolation.

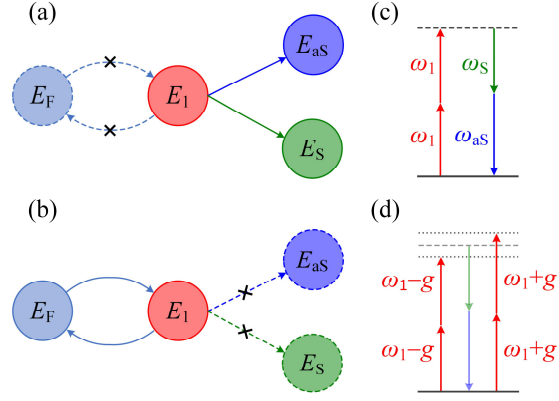

Figure 1. Schematic of the nonreciprocal SPFWM process by unidirectionally coupling the probe mode $E_1$ to an auxiliary vacuum mode $E_F$. (a) and (b) Schematic of the probe-vacuum coupling and the SPFWM in the backward and forward cases, respectively. (c) and (d) Corresponding virtual photon processes. In the backward case (a) and (c), energy conservation is met and the SPFWM occurs when $E_1$ decouples from $E_F$. In the forward case (b) and (d), $E_1$ is dressed by the strong coupling (under a strength of $g$) with $E_F$, and the SPFWM process is suppressed due to the breaking of energy conservation.

Figure 1 schematically shows the key idea for nonreciprocal SPFWM process induced by a direction-dependent probe-vacuum interaction in a nonlinear medium. In a medium with the third-order ($\chi^{(3)}$) nonlinearity, a probe field mode $E_1$ (with frequency $\omega_1$) can spontaneously convert to a pair of Stokes mode $E_S$ and the anti-Stokes mode $E_{aS}$ (with frequencies $\omega_S$ and $\omega_{aS}$), when energy conservation ($2\omega_1=\omega_S+\omega_{aS}$) and PMC hold simultaneously. Conventionally, this spontaneous parametric process is independent of the propagation direction of the probe field, thus reciprocal. Here we show that the inherent reciprocity can be broken when the probe field mode directionally couples to an auxiliary mode $E_F$ [which is from another pumped FWM (PFWM) process driven by two unidirectionally propagating pump laser beams], in combination with a broadband dispersive nonlinear medium. Generally, in the backward case, the probe mode decouples from the auxiliary mode. Consequently, the PMC retains and the SPFWM process occurs efficiently. By contrast, the probe mode strongly couples to the auxiliary vacuum mode $E_F$ in the forward case. This coupling dresses $E_1$ into energies $\omega_1 \pm g$ and breaks the energy conservation condition. As a result, the SPFWM is off. In this sense, we create a nonreciprocal spontaneous parametric process. Below, we demonstrate such chiral probe-vacuum coupling and the enabled broadband optical isolation in Na atoms.

As a proof-of-principle experiment, we realize the SPFWM and PFWM processes in an ensemble of hot Na atoms. The probe field $\boldsymbol{E}_1$ (frequency $\omega_1$, wave vector $\mathbf{k}_1$, Rabi frequency $G_1$, horizontal polarization) and two pump fields $\boldsymbol{E}_2$ and $\boldsymbol{E}_2'$ ($\omega_2$, $\mathbf{k}_2$ and $\mathbf{k}_2'$, $G_2$ and $G_2'$, vertical polarization) from two dye lasers are injected into Na atomic vapors confined in a heating pipe oven with a length of 20 cm. The probe beam is set to co- and counter-propagates with pump beams intersecting with a small angle of ~0.3° inside the medium. See more details of experimental setup in Fig. S1 in the Supplementary Materials.

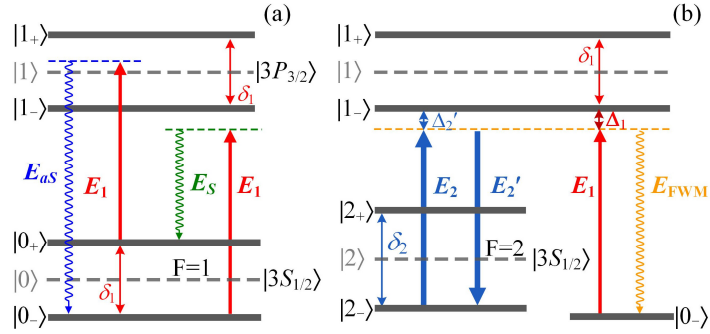

Figure 2. (a) Double-Λ atomic energy-level configuration for the SPFWM. The probe detuning $\Delta_1=\omega_{10}-\omega_1$ is defined as the transition frequency of the two original energy levels [the ground state $|3S_{1/2}, F=1\rangle$ ($|0\rangle$) and the excited state $|3P_{3/2}\rangle$ ($|1\rangle$)] and the frequency of $\boldsymbol{E}_1$. The states $|0_+\rangle$ and $|0_-\rangle$ ($|1_+\rangle$ and $|1_-\rangle$) with a frequency difference of $\delta_1$ represent two dressed states of the ground level $|0\rangle$ (the excited level $|1\rangle$) induced by $\boldsymbol{E}_1$. (b) Three-level atomic energy-level configuration for the PFWM. The states $|2_+\rangle$ and $|2_-\rangle$ with frequency difference of $\delta_2$ are two dressed states of another ground level $|2\rangle$ ($|3S_{1/2}, F=2\rangle$) induced by pump beams. The pump detuning is defined as $\Delta_2=\omega_{12}-\omega_2$.

The energy-level configuration for the SPFWM is depicted in Fig. 2(a). The probe field $\boldsymbol{E}_1$ drives the transition $|3S_{1/2}\rangle\rightarrow|3P_{3/2}\rangle$ and enables the SPFWM process to generate the Stokes mode $E_S$ and the anti-Stokes mode $E_{aS}$ via virtual photon virtual photon process. The "double-Λ" atomic configuration involves the dressing states $|0_\pm\rangle$ of the ground state $|3S_{1/2}, F=1\rangle$ ($|0\rangle$) and $|1_\pm\rangle$ of the excited state $|3P_{3/2}\rangle$ ($|1\rangle$). According to the dressing-state picture [61], the probe beam can simultaneously split states $|0\rangle$ and $|1\rangle$ into new virtual states $|0_\pm\rangle$ and $|1_\pm\rangle$, respectively, with the same frequency difference of $\delta_1=\lambda_{0+}-\lambda_{0-}=(\Delta_1^2+G_1^2)^{1/2}$, where $\lambda_{0\pm}(=\lambda_{1\mp})=\Delta_1/2\pm(\Delta_1^2+G_1^2)^{1/2}/2$ are the eigenvalues of states $|0_\pm\rangle$ ($|1_\mp\rangle$) [62,63]. The frequencies of the conjugated modes $E_{aS}$ and $E_S$ are $\omega_1\pm\delta_1$, respectively, indicating energy conservation. We define the detunings as $\Delta_{aS}=\omega_{10}-\omega_{aS}=\Delta_1-\delta_1$ and $\Delta_S=\omega_{10}-\omega_S=\Delta_1+\delta_1$. When only the probe field propagates in the atomic vapor (the probe-only case), the SPFWM occurs with the same efficiency for the opposite-input probe fields, thus is reciprocal. This reciprocity can be broken when the probe field interacts with another field in a chiral way.

In the level configuration shown in Fig. 2(b), the probe and pump fields excite a PFWM process with an auxiliary vacuum mode $E_F$ with wave vector $\mathbf{k}_F$ and frequency $\omega_F$, mediated by virtual photon virtual photon process. Here the pump beams also create two dressing states $|2_\pm\rangle$ with corresponding eigenvalues being $\lambda_{2\pm}=\Delta_2/2\pm(\Delta_2^2+G_2^2+G_2'^2)^{1/2}/2$. The states $|2_+\rangle$ and $|2_-\rangle$ separate by energy of $\delta_2=\lambda_{2+}-\lambda_{2-}=(\Delta_2^2+G_2^2+G_2'^2)^{1/2}/2$. When the probe and pump fields propagate in almost opposite directions (the forward case), the PMC ($\mathbf{k}_F=\mathbf{k}_1+\mathbf{k}_2-\mathbf{k}_2'$) and energy conservation ($\omega_F=\omega_1+\omega_2-\omega_2'$) are satisfied simultaneously to drive the PFWM. The two-photon resonant condition for the PFWM is $\Delta_1-\Delta_2'=0$ with $\Delta_2'=\Delta_2-\lambda_{1+}+\lambda_{2-}$, by considering different energies for the dressing states caused by the probe and pump beams. As a result, the probe field strongly interacts with the mode $E_F$ in this counter-propagation case and is dressed into different energies. This breaks the condition for energy conservation in the SPFWM process under the same PMC, i.e. $2\mathbf{k}_1=\mathbf{k}_S+\mathbf{k}_{aS}$. As a strong contrast, the PFWM process is off due to the phase mismatching when the probe field co-propagates (the backward case) at a small angle with the strong pump laser beams. In this case, the probe field decouples from the mode $E_F$. Thus, it converts to the Stokes and anti-Stokes modes via the SPFWM with a high efficiency, displaying a strong loss. With the chiral level configurations in Fig. 2, we create a nonreciprocal SPFWM process and can further achieve nonreciprocal transmission of the probe laser beam.

The evolutions of the probe field, the generated Stokes and anti-Stokes fields (SPFWM signal), and the auxiliary field (PFWM signal) are critically dependent on the atomic linear and third-order nonlinear susceptibilities. We assume the atomic number density $N$ and consider the transition between the states $|m\rangle$ and $|n\rangle$ ($m$, $n$=0, 1, 2). $\Gamma_{mn}$

is the corresponding decoherence rate, and $\mu_{mn}$ is the electric dipole moment of the transition. The Rabi frequency of the field with a complex amplitude $E_i$ is calculated as $G_i=\mu_{mn}E_i/\hbar$. The linear susceptibility is given by $\chi_i^{(1)}=N\mu_{mn}^2\rho_i(\omega_i)/(\varepsilon_0\hbar G_i)$, where $\rho_i(\omega_i)=iG_i/(\Gamma_{mn}+i\Delta_i)$ is the first-order density matrix element according to the density matrix method [64], Then, The linear absorption coefficient is evaluated as $\gamma_i=(\omega_i/c)\mathrm{Im}\chi_i^{(1)}$. It determines the decay of the field during propagation in atomic vapors.

We calculate the corresponding third-order density matrix elements $\rho_i^{(3)}$ with the same method as the first-order elements $\rho_i$. In the dressing-state presentation, $\rho_S^{(3)}$, $\rho_{aS}^{(3)}$ and $\rho_F^{(3)}$ are given by:

$$\rho_s^{(3)} = \frac{-iG_1^2 G_{aS}^*}{(\Gamma_{10E}+i\Delta_1+G_F^2/\Gamma_{00E})[\Gamma_{00E}+i\delta_1][\Gamma_{10E}+i(\Delta_1+\delta_1)+G_F^2/(\Gamma_{00E}+i\delta_1)]}, \quad (1.1)$$

$$\rho_{as}^{(3)} = \frac{-iG_1^2 G_S^*}{(\Gamma_{10E}+i\Delta_1+G_F^2/\Gamma_{00E})(\Gamma_{00E}-i\delta_1)[\Gamma_{10E}+i(\Delta_1-\delta_1)+G_F^2/(\Gamma_{00E}-i\delta_1)]}, \quad (1.2)$$

$$\rho_F^{(3)} = -iG_1|G_2|^2 \Big/ \big[(\Gamma_{10E}+i\Delta_1)(\Gamma_{21E}+i\Delta_2')[\Gamma_{20E}+i(\Delta_1-\Delta_2')]\big]. \quad (1.3)$$

The corresponding nonlinear susceptibilities for the SPFWM and PFWM processes are $\chi_{aS/S}^{(3)}=|N\mu_{10}\rho_{aS/S}^{(3)}/(\varepsilon_0 E_1^2 E_{S/aS})|$ and $\chi_F^{(3)}=|N\mu_{10}\rho_F^{(3)}/(\varepsilon_0 E_1 E_2^2)|$, respectively.

The natural decoherence rate $\Gamma_{mn}$ for single Na atom without interacting with light is about tens of megahertz. For thermal atomic ensembles adopted in our experiment, the Doppler and power-induced broadening of the decoherence rate need to be considered. The effective decoherence rate $\Gamma_{mnE}$ in the presence of the probe field becomes $\Gamma_{mnE}=\Gamma_D+\Gamma_P$. For a given Rabi frequency $G_i$, the power broadening can be estimated as $\Gamma_P=\Gamma_{mn}(1+G_i^2/\Gamma_{mn}^2)^{1/2}$ [65]. The Doppler broadening is about 11 GHz calculated by $\Gamma_D=2(\ln2)^{1/2}u\omega_i/c$ with $u=(2K_B T/w)^{1/2}$, where $w$ is the mass of a single Na atom, $K_B$ is the Boltzmann's constant, and $T$ is the absolute temperature. When only the probe beam is on with a Rabi frequency of $2\pi\times6$ GHz, the effective $\Gamma_{10E}$ (for simulating SPFWM) is about 48.7 GHz ($\Gamma_P$=37.7 GHz, $\Gamma_{10}$ =61.5 MHz). The power and Doppler broadening effects together make great contributions to expanding the valid range of bandwidth for the nonreciprocal spontaneous parametric process.

The SPFWM and PFWM processes is governed by the following effective interaction Hamiltonian between field modes [64]:

$$H_I = i\hbar[(\kappa_S + \kappa_{aS})/2]\hat{a}_1^2 \hat{a}_S^\dagger \hat{a}_{aS}^\dagger + i\hbar\kappa_F \hat{a}_2 \hat{a}_2'^\dagger \hat{a}_F^\dagger \hat{a}_1 + H.c. \quad (2)$$

Where $\hat{a}_i$ ($i=aS, S, 1, 2, F$) is the bosonic annihilation operator for corresponding fields, the annihilation operator $\hat{a}_2'$ is for the $\boldsymbol{E}_2'$ field, the coefficient $\kappa_i = |-ik_i\chi_i^{(3)}|$ represents the coupling strength depending on the third-order nonlinear susceptibility $\chi_i^{(3)}$. In the limitation of the strong classical fields, we can replace the annihilation operator $\hat{a}_i$ with the complex amplitude $E_i$ of the positive-frequency component of the corresponding field.

From the nonlinear Helmholtz equation under the slowly-varying amplitude approximation [53,64], we obtain the coupled-mode equations describing the SPFWM and PFWM processes as:

$$\partial E_1/\partial z = -\gamma_1 E_1 - (\kappa_S + \kappa_{aS})E_1^* E_S E_{aS} \times M - \kappa_F |E_2|^2 E_F, \quad (3.1)$$

$$\partial E_s/\partial z = -\gamma_S E_S + \kappa_{aS} |E_1|^2 E_{aS}^* \times M + \beta_S, \quad (3.2)$$

$$\partial E_{aS}/\partial z = -\gamma_{aS} E_{aS} + \kappa_S |E_1|^2 E_S^* \times M + \beta_{aS}, \quad (3.3)$$

$$\partial E_F/\partial z = -\gamma_F E_F + \kappa_F |E_2|^2 E_1. \quad (3.4)$$

$\beta_i = \sqrt{2\gamma_i}\hat{\xi}_i(z)$ is the Langevin noise modeling the quantum fluctuation [66-68], which is important for the spontaneous parametric process. The noise operator satisfies the correlation functions $\langle\hat{\xi}_i^\dagger(z)\hat{\xi}_i(z')\rangle = 0$, $\langle\hat{\xi}_i(z)\hat{\xi}_i^\dagger(z')\rangle = \delta(z-z')$. Here, $\kappa_F=0$ is for the backward case while the nonzero $\kappa_F$ for the forward case. Because the pump field $\boldsymbol{E}_2$ and $\boldsymbol{E}_2'$ are strong and propagate in almost the same direction, we consider them constant during propagation. For simplicity, we replace $E_2'$ with $E_2$ due to their identical intensity. In experiment, the Stokes and anti-Stokes fields can be generated in pairs in the SPFWM over a wide range of the frequency and wave vectors, exhibiting a cone in the momentum space around the probe beam [69,70], and contains hundreds paired spatial modes [71]. Each pair of Stokes and anti-Stokes modes is defined as $E_S$ and $E_S'$ in the theoretical model in Eq. (3). To consider this effect, we multiple the SPFWM interaction terms by a factor M.

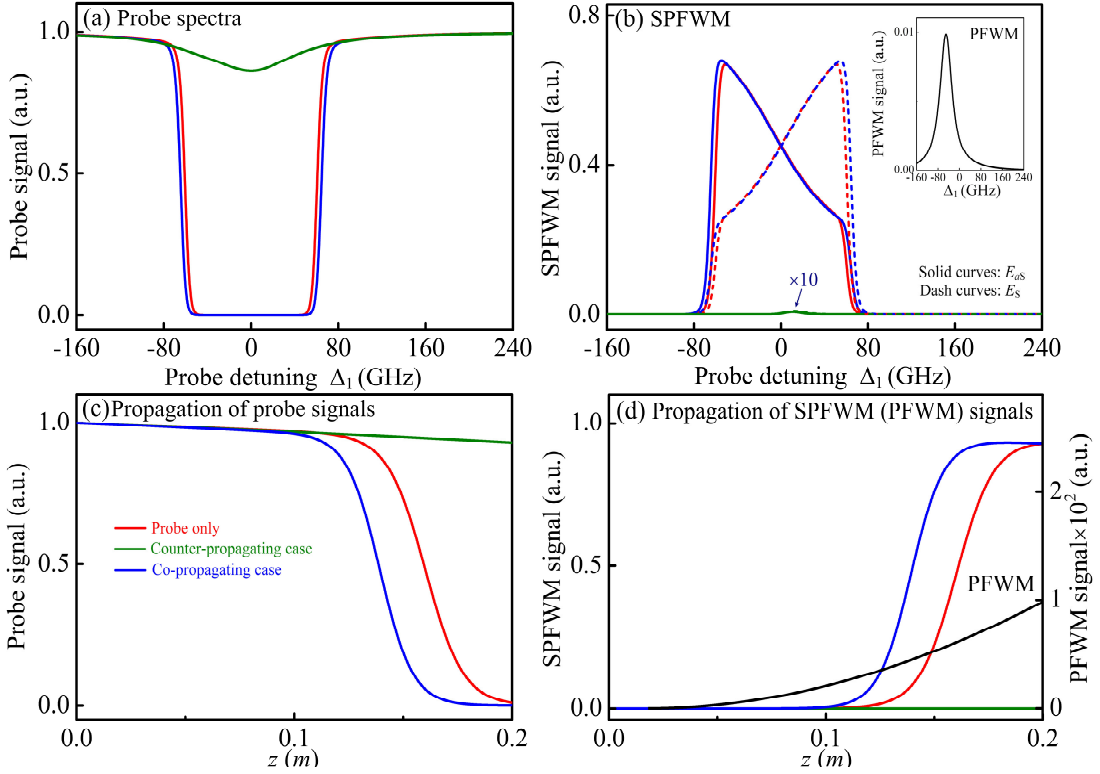

Figure 3. Simulated output spectra and evolution of the probe field, the SPFWM and PFWM signals in the atomic vapors in the cases of probe only (red), forward (green), and backward (blue) cases. All intensities are normalized by the input intensity of the probe laser beam. (a) Spectra of the probe field. (b) Spectra of the SPFWM and PFWM (inset) signals at $\Delta_2'=-2\pi\times8.2$ GHz. (c) Evolution of the probe field. (d) Evolution of the SPFWM and PFWM (black curve) signals at $\Delta_1=\Delta_2'=-2\pi\times8.2$ GHz. The propagation of PFWM signal is also provided for the forward case. Other parameters are: $N=1.24\times10^{14}$ cm$^{-3}$ (the forward case) and $1.43\times10^{14}$ cm$^{-3}$ (the backward case), $G_1=2\pi\times6$ GHz, $G_2=G_2'=2\pi\times30$ GHz, and M=600.

Figures 3(a) and (b) show the simulated transmitted spectra of the probe field, the SPFWM and PFWM signals after passing the atomic vapor for three cases of probe only, the forward case (the counter-propagating case) and the backward case (the co-propagating case). The SPFWM process is strong in the probe-only and backward cases. As a result, the probe field is completely absorbed over a broad bandwidth [see red and blue curves in Fig. 3(a)] due to the resonant absorption and the SPFWM. The generated SPFWM signals [red and blude curves in Fig. 3(b)] are similar in both cases. Because the anti-Stokes and Stokes modes are different in frequency, the Stokes and anti-Stokes modes are symmetric with respect to $\Delta_1=0$. In the backward case, the PFWM process is weak because the PMC breaks, and the generated auxiliary field is negligible. Consequently, the probe field in both cases first decays slowly when the Stokes and anti-Stokes are weak. Then, it decays from $z=0.13$ m ($z=0.12$ m) rapidly to vanishing small at $z=0.2$ m ($z=0.16$ m) in the probe-only (backward) case, see Fig. 3(c). Correspondingly, the generated SPFWM signals increase fast to a saturated intensity during this propagation distance. As shown in Figs. 3(a) and (b), the probe transmission is lower and the SPFWM signals is stronger in the backward case than the probe-only case. This is because the optical pumping effect of the pump fields increases the effective atomic density [56,72], leading to a stronger SPFWM process.

In stark contrast, the forward case shows a strong nonreciprocal transmission. The PFWM process occurs efficiently because the PMC and energy conservation condition meet simultaneously. This is reflected by the strong PFWM signals (black curves) shown in Figs. 3(b) and (d). Because the pump field is strong, this PFMW process create a strong interaction between the weak probe field and the auxiliary vacuum mode $E_F$. This chiral interaction dresses the energy of the probe field and unidirectionally breaks the energy conservation condition for the SPFWM process. Therefore, the SPFWM is greatly suppressed and the probe field can transmit through the atomic vapor with low absorption, see Figs. 3(a) and (c), indicated by the very weak SPFWM signal (green curves) in Fig. 3(d). The nonreciprocal SPFWM and probe transmission are in coincidence with the calculated third-order susceptibilities for three cases (see in Fig. S2 in Supplementary Material).

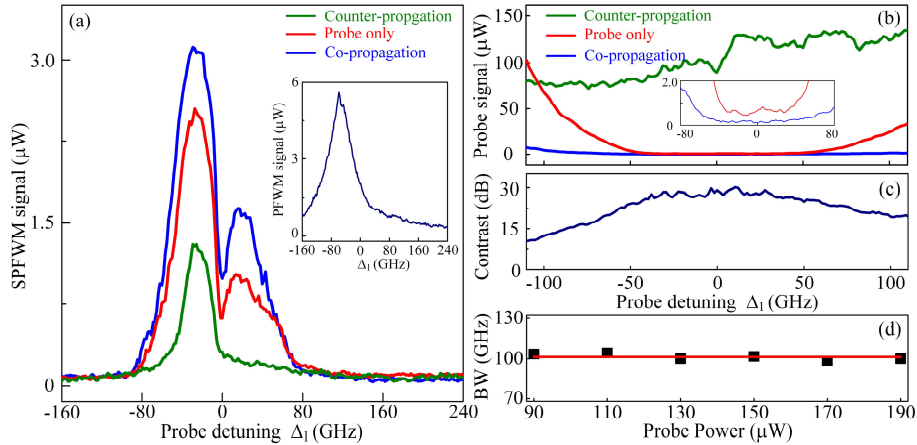

Figure 4. (a) Measured SPFWM spectra and (b) transmitted probe spectra as a function of the probe detuning for only a unidirectional probe beam applied (red), the co-propagation case (blue) and counter-propagation case (green). The inset in (a) shows the observed PFWM spectrum. (c) Nonreciprocal transmission contrast calculated from (b). (d) Bandwidth (BW) for isolation ratio >25 dB versus the probe power when two probe beams propagate simultaneously in opposite directions. The square are experimental data, while the solid line represents a fitting and provides a guide to the eye. The probe power is 150 μW in (a-c). The power of the two pump beams are 3.75 mW.

The nonreciprocity of the SPFWM is experimentally displayed in Fig. 4(a). Considering the spatial ring structure of the SPFWM process, the anti-Stokes and Stokes modes are undistinguishable in space [65]. The captured SPFWM spectra contains both the anti-Stokes and Stokes modes, and exhibits a double-peak profile, corresponding to the sum of anti-Stokes and Stokes components (see theoretical Fig. S3 in the Supplementary Material). Here we measure the power spectra of the Stokes and anti-Stokes modes generated in the SPFWM process in three cases: (i) in the absence of the pump laser beams (red curve); (ii) the forward case (green curve); (iii) the backward case (blue curve). The SPFWM processes in the cases (i) and (iii) are strong because the PFWM process is prevented and the probe mode decouples from the auxiliary mode. As a result, the probe field converts to the Stokes and anti-Stokes modes efficiently, see red and blue curves in Fig. 4(a), and the corresponding probe transmissions are weak [see Fig. 4(b)]. Because the strong pump fields repump the atoms, the conversion is stronger and the transmission is lower in the backward case. In distinct contrast, the PFWM happens in the forward case. It creates a strong interaction between the probe and auxiliary modes. This interaction dresses the probe mode to break the PMC for SPFWM process over a wide frequency range. Therefore, the SPFWM signal is remarkably suppressed [green curve in Fig. 4(a)] and the corresponding probe transmission is high [green curve in Fig. 4(b)].

The intensity of the observed SPFWM power spectra exhibits an asymmetric structure with respect to $\Delta_1=0$, and this can be attributed to the other transition $|3S_{1/2}\rangle \rightarrow |3P_{1/2}\rangle$ of sodium atoms. Considering the Doppler and power-induced broadening effect on the $|3P_{1/2}\rangle$, the incident probe beam (resonant with $|3S_{1/2}\rangle \rightarrow |3P_{3/2}\rangle$) can simultaneously excite the transition $|3S_{1/2}\rangle \rightarrow |3P_{1/2}\rangle$, but under the far-detuning

condition. Namely, both transitions can produce absorption for the probe beam and the generated SPFWM signals. A smaller $\Delta_1$ means that frequency of probe field is set further from the resonance of state $|3P_{1/2}\rangle$, which hence gives rise to a weaker absorption. The observed asymmetric intensity profile of the probe transmission (probe only case) advocates this effect (see Fig. S4 in the Supplementary Material). When the absorption from the transition $3S_{1/2}\rightarrow 3P_{1/2}$ is theoretically considered, the simulated SPFWM spectra exhibits an asymmetric double-peak profile (see Fig. S3 in the Supplementary Material).

The transmitted probe intensities of the forward and backward cases are denoted as $T_f$ and $T_b$, and we calculate the isolation ratio as $\eta=10\log_{10}(T_f/T_b)$. The contrast dependence on the probe detuning is presented in Fig. 4(c) with either probe laser beam selectively on. The contrast is larger than 25 dB over 100 GHz band (ranging from approximately −45 GHz to 55 GHz) and reaches the maximum value 30 dB at $\Delta_1\approx 10$ GHz.

Our nonreciprocal regime can bypass dynamic reciprocity [73], which impose strong constraint on most nonlinear isolators, when two probe laser beams are oppositely incident to the medium at the same time. The transmission contrast versus the probe detuning is close to the single-probe contrast shown in Fig. 4(c), see Fig. S5 in the Supplementary Material. The nonreciprocal bandwidth for contrast >25 dB exceeds 100 GHz and remains stable as the input power increases from 90 μW to 190 μW [see Fig. 4(d)]. This result indicates that the demonstrated optical isolation exhibits robustness against the input probe power over a broad bandwidth.

In summary, we experimentally demonstrated a non-reciprocal spontaneous parametric process of a weak probe field by breaking its inversion symmetry and time reversal symmetry by introducing a strong-field-driven PFWM. Based on this nonreciprocal process, we achieved optical isolation with large isolation ratio and broad bandwidth. Spontaneous parametric processes and PFWM processes have been realized in diverse nonlinear platforms with chip-compatible materials such as $LiNbO_3$ and silicon. The concept of this work paves the way for integrated optical isolation by means of nonlinear wave conversion and thus can boost the functions of photonic chips.

This work was supported by National Natural Science Foundation of China (Grant Nos. 62022066, 12074306 and 92365107), the National Key R&D Program of China (Grant No. 2019YFA0308700), Key Scientific and Technological Innovation Team of Shaanxi Province (2021TD-56), the Program for Innovative Talents and Teams in Jiangsu (Grant No. JSSCTD202138).